\documentclass[10pt]{article}
\usepackage[letterpaper]{geometry}
\usepackage{hicss}
\usepackage{times}
\usepackage[none]{hyphenat}
\usepackage{url}
\usepackage{latexsym}
\usepackage[frozencache=true]{minted}
\usepackage{indentfirst}
\usepackage{graphicx}
\graphicspath{{images/}}
\usepackage[style=apa]{biblatex}
\usepackage{hyperref}
\usepackage{amsmath,amssymb}
\usepackage[capitalise]{cleveref}
\usepackage{enumitem}
\usepackage{ifthen}
\newboolean{showcomments}
\setboolean{showcomments}{true} 
\ifthenelse{\boolean{showcomments}}
  {\newcommand{\nb}[2]{
    \fcolorbox{gray}{yellow}{\bfseries\sffamily\scriptsize#1}
    {\sf\small$\blacktriangleright$\textit{#2}$\blacktriangleleft$}
   }
   
  }
  {\newcommand{\nb}[2]{}
   
  }

\title{CROSS: A Contributor-Project Interaction Lifecycle Model for Open Source Software}

\author{Tapajit Dey \\
 Lero--SFI Software Research Centre \\
 University of Limerick \\
 {\underline{tapajitdey@acm.org}} \\ \\
 \And
 Brian Fitzgerald \\
 Lero--SFI Software Research Centre \\
 University of Limerick \\
 {\underline{brian.fitzgerald@ul.ie} } \\ \\
 \And
 Sherae Daniel \\
 University of Cincinnati \\
 {\underline{daniesr@ucmail.uc.edu} } \\ \\
 }

\date{}

\begin{document}
\maketitle
\begin{abstract}
Despite the widespread adoption of open source software (OSS), its sustainability remains a critical concern, particularly in light of security vulnerabilities and the often inadequate end-of-service (EoS) processes for OSS projects as they decline. Existing models of OSS community participation, like the Onion model and the episodic contribution model, offer valuable insights but are fundamentally incompatible and fail to provide a comprehensive picture of contributor engagement with OSS projects. This paper addresses these gaps by proposing the CROSS model, a novel contributor-project interaction lifecycle model for open source, which delineates the various lifecycle stages of contributor-project interaction, along with the driving and retaining forces pertinent to each stage. By synthesizing existing research on OSS communities, organizational behavior, and human resource development, it explains a range of archetypal cases of contributor engagement and highlights research gaps, especially in EoS/offboarding scenarios. The CROSS model provides a foundation for understanding and enhancing the sustainability of OSS projects, offering a robust foundation for future research and practical application.
\end{abstract}

\subsubsection*{Keywords:} 

open source software, onion model, onboarding, offboarding, life-cycle model 

\section{Introduction}
Open source software (OSS) has become a critical piece of modern digital infrastructure, with almost all industries (including many non-IT companies) relying on OSS in both the public and private sectors. Although the history of open source software dates back decades, with the launching of the GNU project in 1983 and the founding of the Free Software Foundation in 1985, its popularity has skyrocketed in the last couple of decades. As of June 2024, GitHub~\footnote{\url{https://github.com/about} --- Accessed on 2024-06-03)} is reported to have over 100 million developers and over 420 million repositories. Open source ecosystems have emerged to effectively manage the vast number of open source projects, with private companies and volunteer communities working side-by-side [\cite{franco2017open}]. [~\cite{steinmacher2017free}] have termed this proliferation and maturation in open source as ``the end of the teenage years", given the timeline since the open source term was first coined. Likewise, the roles [\cite{canovas2022analysis}], motivations [\cite{wang2022influence,shimada2022github}], work practices  [\cite{drost2021open,hmoud2022open}], and nature \footnote{\url{https://www.statista.com/chart/25795/active-github-contributors-by-employer/} --- Accessed on 2024-06-03)} of open source contributors have also changed significantly since the early days of OSS development. 

However, even with Big Tech companies (e.g., Google, Microsoft, Amazon, etc.) contributing significantly to various open source projects in the past few years, open source software development largely depends on volunteers across the world [\cite{o2021coproduction}]. Thus, maintaining a steady influx of new developers [\cite{alexander2002working}] and retaining developers for the long term [\cite{Yamashita2016}] are critical for the survival and continuing success of an open source project, the community at large, and indeed of the overall software industry ecosystem. 

Given the volunteer-driven nature of OSS and its sustainability challenges, it's essential to understand the dynamic interactions between contributors and projects to identify pain points and research opportunities. Here, we use the term ``contributors'' instead of ``developers'' to encompass those who support OSS projects in non-coding roles, e.g., by providing user support or writing documentation, that is vital to the community [\cite{jin2007beyond}].

Previous studies have examined various phases of interaction between open source projects and their contributors. For instance, some research focused on the initial stages of contributor engagement, particularly the joining phase, where new contributors become acquainted with a project and begin to participate actively e.g., [\cite{Robles2006, VonKrogh2003}]. Other studies, such as [\cite{crowston2005social,Sinha2011}], have explored the progression of developers within OSS communities, shedding light on how contributors advance to more central roles over time. 
However, the existing literature primarily focuses on specific segments of the contributor journey, such as onboarding or role progression, without providing a comprehensive view of the entire lifecycle of contributor engagement with OSS projects. As a result, critical periods of contributor-project interaction, such as offboarding and long-term disengagement, remain underexplored. 
This oversight can lead to several significant challenges. In OSS communities, the departure of key contributors can result in substantial knowledge loss [\cite{rashid2017exploring}], disruptions in development, and even project abandonment [\cite{avelino2019abandonment}]. Unlike traditional organizations with formalized processes and resources for managing employee transitions, OSS projects often lack structured practices for managing contributor transitions. The absence of formal offboarding protocols can lead to the loss of valuable insights, expertise, and historical knowledge, creating gaps that remaining contributors may struggle to fill. 

Addressing these gaps is crucial for developing a comprehensive understanding of OSS contributor dynamics, improving project sustainability, and ensuring the long-term health and security of OSS communities. Structured research on offboarding practices and the effects of long-term disengagement of key contributors can provide essential insights into maintaining project continuity, preserving critical knowledge, and developing strategies to re-engage valuable contributors, thereby enhancing the overall sustainability and resilience of OSS projects.
To address these issues, we introduce the \textbf{Contributor Project Interaction Lifecycle for OSS (CROSS) Model}, which comprehensively captures the full lifecycle of contributor interaction with OSS projects. Thus, the \textbf{CROSS} model not only places existing research within a broader framework but also identifies critical phases and transitions that have been overlooked, offering new avenues for future research and practical applications in sustaining OSS communities.

This paper makes the following key contributions:

\noindent\textbf{--} We introduce the Contributor Project Interaction Lifecycle for OSS (\textbf{CROSS}) Model, the first comprehensive lifecycle model that captures the full range of contributor interactions with OSS projects.\\
\noindent\textbf{--} We discuss the significance of the often-overlooked offboarding phase, highlighting its critical role in maintaining project continuity and mitigating security risks within OSS communities. \\
\noindent\textbf{--} We provide a unified explanation of various types of contributors, such as episodic volunteers, one-time contributors, and core developers, and their different journeys through OSS projects, as illustrated in \Cref{fig:dev}. This offers a more complete understanding of contributor dynamics within a single, cohesive model.



\section{Model Development}

\subsection{Background and Related Work}

While there is no comprehensive model describing the OSS contributor-project interaction lifecycle, studies have focused on specific phases of the interaction. [\cite{steinmacher2014attracting}] outlined a ``joining model'' for OSS developers and depicted the stages of joining and forces that act during the joining process. 
The ``Onion model'' describing a contributor's progression through an OSS project/community has gained prominence [\cite{crowston2005social}]. This model identifies four layers of participants. At the center of the onion are the core developers who contribute most of the code and oversee the design and evolution of the project. In the next ring are the co-developers who submit patches that are reviewed and checked in by core developers. Further out are the active users who do not contribute code but provide use–cases and bug reports as well as test new releases. Further out still are the passive users of the software who do not contribute to the project channels. The Onion model has gained widespread acceptance and remains one of the most influential models for describing OSS communities.

However, while the Onion model captures the state of the community around an OSS project, it does not adequately capture an individual contributor's journey through an OSS community. For instance, it has been questioned in terms of its assumption that participants progress through the various layers [\cite{jergensen2011onion}]. Research on episodic volunteering has found that some contributors do not seek to become habitual contributors nor to progress in the hierarchy [\cite{barcomb2018uncovering}]. Moreover, the Onion model only focuses on developer progression through OSS communities and does not capture the developer churn phenomenon. Therefore, in this research, as an alternative to the Onion model, we propose the \textbf{CROSS} model which focuses on the complete lifecycle of a contributor's interaction with an OSS project/community. 

\subsection{Synthesis of the CROSS Model}\label{ss:syn}
The interaction between OSS projects and contributors is inherently complex, with each interaction being unique. However, there are common patterns that can be used to form a conceptual model of project-contributor engagement. While the idea of an OSS contributor lifecycle exists implicitly, and different phases of this lifecycle have been explored in various studies, including but not limited to the ones noted above, it has yet to be formalized into a comprehensive model that captures the full range of interactions. To develop such a model, we began by identifying the distinct phases of interaction between contributors and OSS projects, drawing from existing research and theories that describe various forms of contributor engagement. We synthesized these insights to build a model that describes the complete lifecycle of a contributor's engagement with OSS projects and communities - the \textbf{CROSS} model - with each phase informed by relevant theoretical frameworks.

Organizational Socialization Theory [\cite{van1977toward}] explains how individuals acquire the knowledge, skills, and behaviors needed to integrate into an organization. In the context of OSS, this theory informs the onboarding and pre-onboarding phases, helping us understand how contributors, like new employees, learn the norms and practices of a project community. It also plays a role in understanding role progression, illustrating how contributors can move from peripheral roles to leadership positions as they become more engaged in the project.

The role progression aspect is examined in further detail using Role Theory [\cite{katz2015social}]. This theory delves deeper into how individuals understand and perform different roles within an organization and is key to explaining role transitions in OSS projects (similar to the Onion model), such as how contributors move between roles like newcomers, active contributors, and core maintainers. It also informs the offboarding phase of contributor-project interaction, highlighting how contributors transition out of these roles by transferring responsibilities to others and gradually disengaging from the project.

The possible reasons for contributor disengagement can be explained by the Unfolding Model of Voluntary Employee Turnover [\cite{lee1994alternative}]. It emphasizes the role of \textit{shocks} or critical events in a person's decision to leave an organization. This theory also informs the offboarding process in OSS projects by explaining how contributors disengage, whether due to personal circumstances, dissatisfaction, or external opportunities. Understanding this phase is crucial for maintaining project continuity and community stability.

Finally, Human Resource Development (HRD) frameworks [\cite{swanson2022foundations}] provide a comprehensive view of organizational practices such as orientation, training, performance management, and knowledge transfer. These frameworks inform the entire OSS contributor-project interaction lifecycle, including onboarding, offboarding, and the management of contributor transitions across different roles. They highlight the importance of knowledge transfer and exit planning, ensuring that contributors' departures do not disrupt the project's continuity—an aspect often overlooked in existing OSS literature. 

By synthesizing these theoretical insights, we develop the \textbf{C}ontributor P\textbf{r}oject Interaction Lifecycle for \textbf{OSS} Model, or the \textbf{CROSS} model, which captures the full lifecycle of contributor engagement in OSS projects, spanning onboarding, role progression, offboarding, and long-term disengagement. This comprehensive framework addresses gaps in existing models, offering a complete understanding of contributor dynamics in OSS communities.

\subsection{The CROSS Model}
\begin{figure*}
    \centering
    \includegraphics[width=0.8\linewidth]{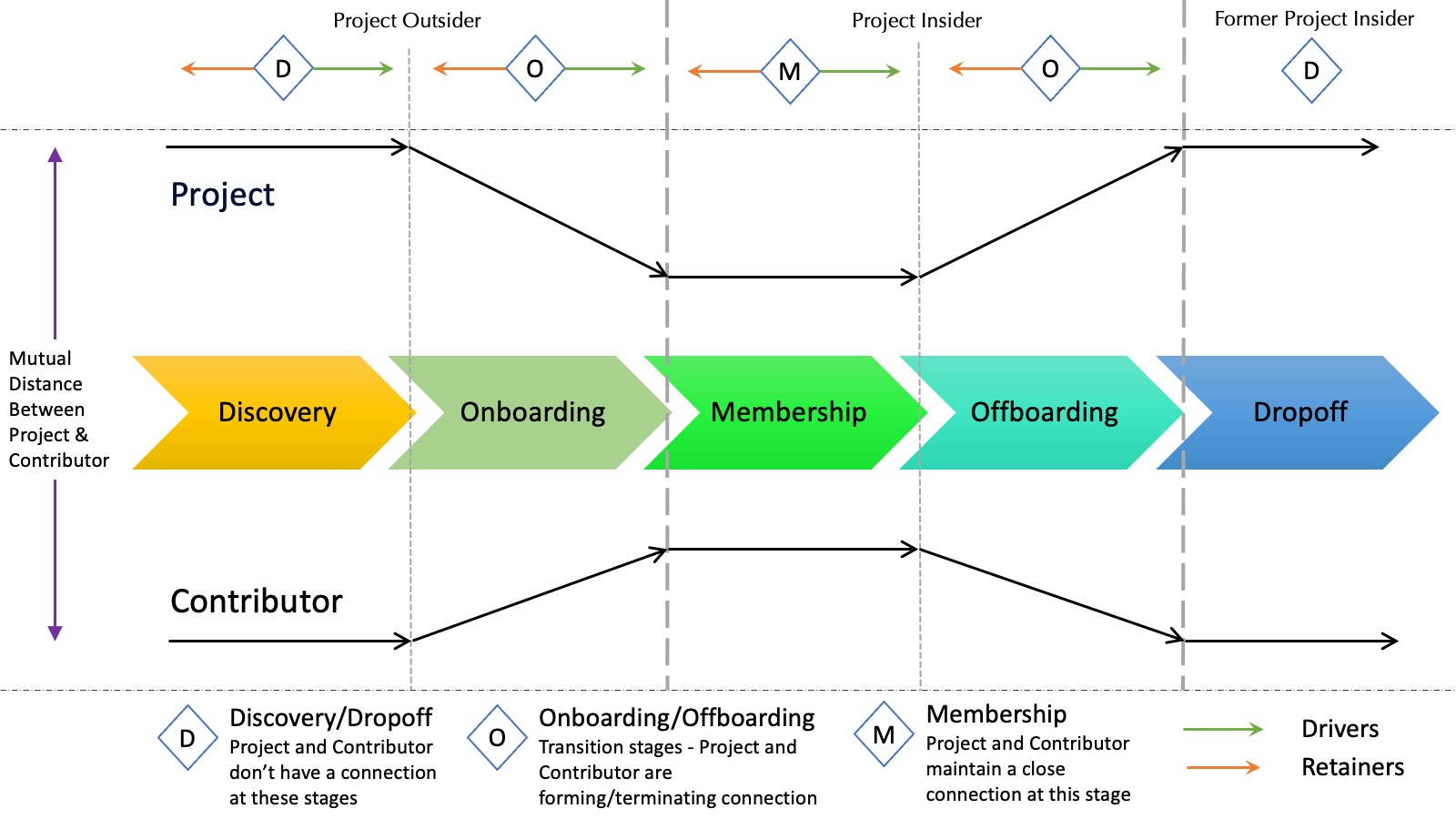}
    \caption{Contributor Project Interaction Lifecycle for OSS (CROSS) Model}
    \label{fig:cross}
    \vspace{-10pt}
\end{figure*}

\Cref{fig:cross} shows an overview of the \textbf{CROSS} model, illustrating the distinct phases of interaction between a contributor and an OSS project/community, the forces that act to retain a contributor in their current phase of interaction or drive them to the next phase, and the mutual distance between the project and the contributor, which is a composite notion indicating how involved a contributor is with the regular operations of the project, comprising of how familiar a contributor is with the project's culture, the development team, the project's road-map and priorities, ongoing concerns and issues, etc. and how much influence they have over these decisions.

\subsubsection{Phases:}

According to our proposed \textbf{CROSS} model, the phases of interaction between a particular open source project/community and a contributor can be broadly divided into five phases: \textit{Discovery, Onboarding, Membership, Offboarding, }and\textit{ Dropoff}, as depicted in \Cref{fig:cross}. 
The definitions of these phases are derived based on standard terminology:

\begin{enumerate}[wide=5pt,noitemsep,parsep=0pt,topsep=2pt]
    \item \textit{Discovery:} This is the phase when a contributor starts getting familiarized with the project/community and does not actually make any contributions to the project. This phase starts when the contributor first encounters the project and lasts until they make a decision to contribute to the project. During this phase, the individual familiarizes themselves with the project but does not make any contributions. The contributor primarily remains a passive user, exploring project documentation, joining mailing lists, and reading discussions among project members. Some contributors may engage in discussions as a form of anticipatory socialization [\cite{van1977toward}]. The mutual distance between the project members and the contributor remains high at this stage. 
    \item \textit{Onboarding:} In this phase, the contributor attempts to get their contribution accepted in the project and might transition into becoming a member. This phase starts when the contributor decides to contribute to the project and ends when they have successfully transitioned into the project. During onboarding, the contributor learns the project’s contribution guidelines, prepares a contribution (often in the form of an issue or pull request), engages with project maintainers for feedback, and refines their contribution as needed. They may also interact with project members who mentor them. The mutual distance between the contributor and project members decreases as familiarity grows.
    \item \textit{Membership:} This phase marks the contributor’s active involvement in the project, where they contribute regularly. This is when a contributor transitions from being an outsider to the project to becoming a project insider. Contributors in this phase may be either engaged, taking on regular maintenance tasks such as reviewing pull requests, addressing issues, and participating in making project evolution decisions, or disengaged, participating less frequently. This phase begins once the contributor's initial contributions are accepted and lasts until they decide to stop contributing. Contributors may hold various roles within the project, such as maintainers or core members, and can transition between roles based on community guidelines. The mutual distance between the contributor and project members is at its lowest during this phase but can vary depending on the level of engagement.
    \item \textit{Offboarding:} This phase represents the journey of a contributor transitioning out of a project. This phase begins when the contributor decides to leave the project and continues until they are completely disengaged. During offboarding, the contributor transitions their responsibilities to other active members and ideally engages in knowledge transfer to ensure continuity. Effective knowledge transfer is critical not only for the continuity of the individual project [\cite{avelino2019abandonment}] but also for the broader OSS community, as unmaintained open-source projects can pose significant security risks within software supply chains [\cite{zahan2022weak}].  The mutual distance between the contributor and project members starts to increase as the contributor disengages from their active roles.
    \item \textit{Dropoff:} This phase starts when the contributor ceases all contributions to the project and effectively retires from active involvement. At this stage, the contributor may remain passively engaged, occasionally participating in discussions, but their regular contributions have ended. The mutual distance between the contributor and project members becomes high once again.
\end{enumerate}

\begin{figure}
    \centering
    \includegraphics[width=1\linewidth]{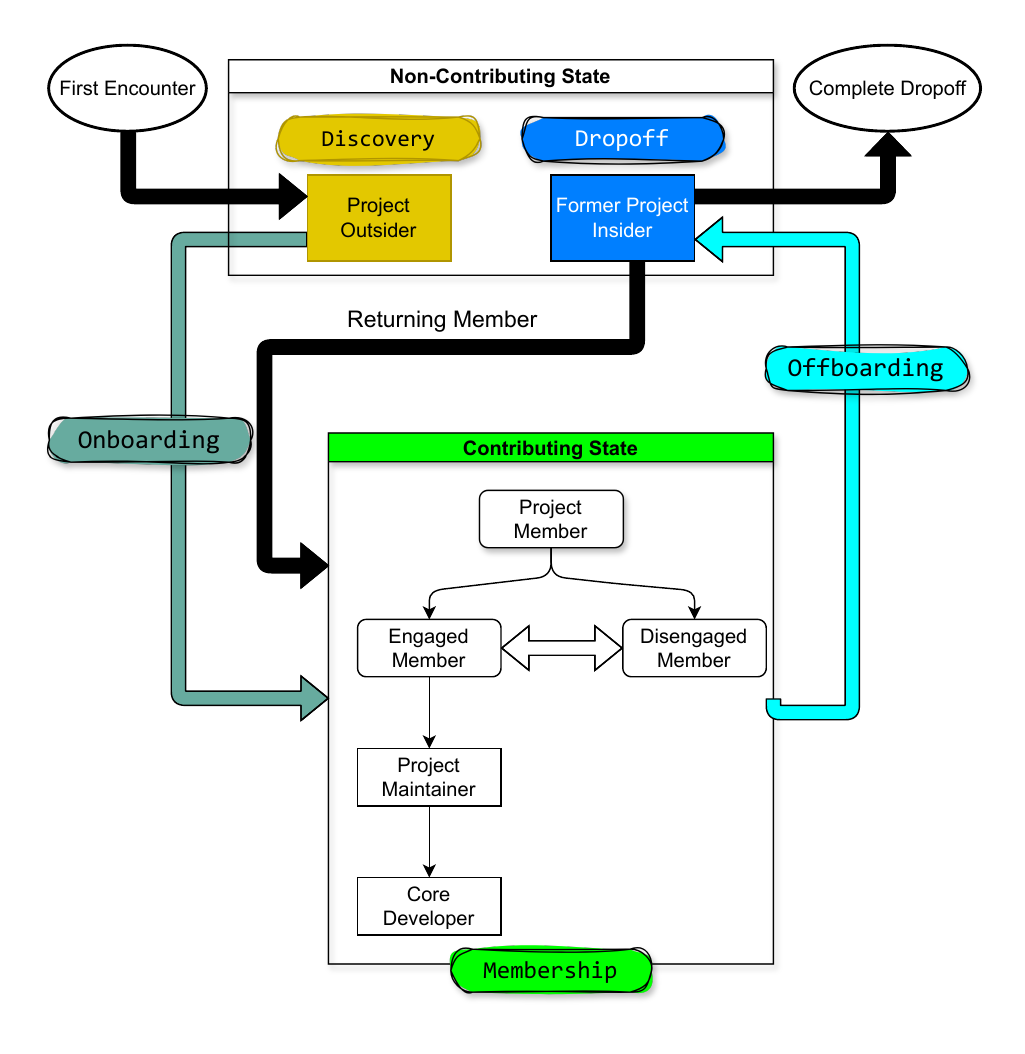}
    \caption{A Contributor's Journey through the Phases of Interaction with an OSS Project -- \\An Alternate View of the \textbf{CROSS} Model}
    \label{fig:dev}
    \vspace{-10pt}
\end{figure}

Before the five phases of the \textbf{CROSS} model, there is technically a pre-discovery phase, during which a potential contributor has no familiarity with a particular OSS project. However, we do not include this phase in the model, as it does not directly impact the contributor-project lifecycle. The phase sequence in the \textbf{CROSS} model is similar to those observed in other contexts, such as an employee in a company or a student at a university. The duration of each phase can vary, and in some cases, a phase may be skipped altogether (e.g., if a project suddenly terminates, cutting off membership). Additionally, not all phases are relevant at any given time for every contributor-project interaction. For instance, while a contributor is actively involved, the Offboarding and Dropoff phases are not applicable. It's also important to note that the Dropoff phase doesn't necessarily signal the end of a contributor's involvement; contributors, such as episodic volunteers, may take breaks and return, a transition clearly illustrated in \Cref{fig:dev}.

One key aspect shown in \Cref{fig:cross} is the distinction between `project insider', `project outsider', and `former project insider'. Contributors in the Discovery or Onboarding phases are considered outsiders, with no prior experience or familiarity with the project's culture. In contrast, those in the Membership, Offboarding, or Dropoff phases have interacted with the project before. This distinction matters because contributors with a history of interaction, even after dropping off, may find it easier to have their future contributions accepted, as demonstrated in prior studies [\cite{dey2020effect}].

\subsubsection{Forces:}

The \textbf{CROSS} model also highlights that there are forces at play in the first four of these five phases that can be broadly categorized into two groups: forces that aid in the transition to the next phase (i.e., \textit{Drivers}) and forces that inhibit such transition and tend to keep the contributors in their current lifecycle phase (i.e., \textit{Retainers}). For the final phase (\textit{Dropoff}), there are no \textit{Drivers} or \textit{Retainers} as it is the final phase in this conceptual model. Some instances of these forces are listed below:

\begin{itemize}[wide=5pt,noitemsep,parsep=2pt,topsep=2pt]
    \item[$\rightarrow$] \textit{Drivers -- Discovery Phase:} The forces that drive a passive user of an OSS project to become an active contributor have been extensively studied in the literature on OSS contributor motivations, which fall broadly into the following categories [\cite{von2012carrots}]: Intrinsic motivators (e.g., Ideology, Kinship, Enjoyment and Fun), Extrinsic motivators (e.g., Pay, Career), and Internalised Extrinsic motivators (e.g., Reputation, Learning, Reciprocity, Own-use). 
    \item[$\leftarrow$] \textit{Retainers -- Discovery Phase:} The forces that prevent a passive user from contributing to an OSS project have also been extensively studied. The literature review by [\cite{steinmacher2015systematic}] highlights a number of such barriers. These include technical challenges, knowledge gaps, a lack of clear documentation, and an unfriendly community environment. There are also certain barriers faced by specific groups of contributors, such as female contributors [\cite{balali2018newcomers}] or contributors from the global south [\cite{prana2021including}], who may encounter cultural biases, lack of support, and access issues. These barriers can significantly hinder the transition from a passive user to an active contributor.
    \item[$\rightarrow$] \textit{Drivers -- Onboarding Phase:} The forces that help contributors successfully complete the onboarding phase and get their contributions accepted are well-documented in various studies. Some of the key drivers include effective mentoring, and providing guidance and support to new contributors [\cite{fagerholm2014role}]. Dedicated onboarding programs, which offer structured pathways for new contributors to learn and engage with the project, also play a significant role [\cite{labuschagne2015onboarding}]. Prior social interaction between project maintainers and contributors can facilitate a smoother onboarding process, as pre-existing relationships can ease communication and collaboration [\cite{tsay2014influence}]. The technical quality and relevance of the submitted contributions are also critical; contributions that meet the project's standards and address pertinent issues are more likely to be accepted [\cite{dey2020effect}]. 
    \item[$\leftarrow$] \textit{Retainers -- Onboarding Phase:} Conversely, the forces that prevent a contributor from successfully completing the onboarding phase are often the opposite of these drivers. A lack of response from project maintainers can leave contributors feeling unsupported and discouraged. Poor or nonexistent mentoring can result in new contributors struggling to navigate the project's requirements and culture. Additionally, a contributor's lack of expertise or failure to adhere to community norms can hinder their contributions from being accepted. These barriers can significantly impede a contributor's progress during the onboarding phase.
    \item[$\rightarrow$] \textit{Drivers -- Membership Phase:} The forces driving a developer to transition out of the membership phase are often similar to those leading them to leave the project. These include the satisfaction of achieving personal or professional goals, experiencing burnout or fatigue, and shifts in personal or professional priorities. Project completion or reaching a stable point where fewer contributions are needed can also prompt disengagement. Additionally, unresolved conflicts, dissatisfaction with the project's direction, or negative experiences within the community may cause contributors to leave. Finally, external opportunities, such as new projects, career advancements, or other personal interests, can draw contributors away from their current OSS commitments [\cite{miller2019people}].
    \item[$\leftarrow$] \textit{Retainers -- Membership Phase:} The forces that influence a contributor’s decision to remain a member of an OSS project and potentially take on higher responsibilities are often similar to their initial motivations for joining. A strong sense of community, built on relationships and bonds within the project, can keep contributors engaged and involved in various discussions and activities. Personal investment in the project also plays a crucial role; contributors who have dedicated significant time and effort may find it difficult to disengage. Established reputation and recognition within the community can further motivate contributors to stay, as they seek to sustain their status and continue being acknowledged for their contributions. Mentorship roles can be another retainer, with experienced contributors finding satisfaction in guiding and supporting new members. Persistent interest in the project's evolution or the technologies it utilizes can keep contributors engaged, even when they are not actively contributing regularly. Additionally, unfinished tasks or pending contributions can retain contributors until they feel their work is fully resolved. Overall, the desire for respect and acceptance, the satisfaction derived from meaningful contributions, and the intrinsic enjoyment of the work can significantly motivate contributors to stay engaged and assume greater responsibilities. These retainers collectively help maintain a stable and active membership base within the OSS project, ensuring its ongoing vitality and success.
    \item[$\rightarrow$] \textit{Drivers -- Offboarding Phase:} The offboarding phase of OSS contributors is not studied very extensively in the literature. However, it is likely that the forces that drive a contributor to complete offboarding and move to the Dropoff phase can be similar to the ones that drive them to leave the project in the first place. These drivers might include personal reasons such as a change of interest, time constraints, or new opportunities elsewhere. Additionally, contributors might feel a sense of completion or fulfillment with their contributions, leading them to move on.
    \item[$\leftarrow$] \textit{Retainers -- Offboarding Phase:} The forces that prevent a contributor from completing their offboarding often stem from the project's needs and dependencies. A lack of existing members to take over their role can be a significant retainer, as the contributor may feel obligated to stay until a suitable replacement is found. Additionally, project members, users, or sponsoring companies (in the case of sponsored OSS projects) might encourage or pressure the contributor to remain involved, recognizing the value and expertise they bring to the project. This sense of responsibility and external pressure can delay the completion of the offboarding process.
\end{itemize}

As for the Dropoff phase, which represents the final stage in the contributor lifecycle where individuals cease all active contributions to the OSS project, there are no specific drivers pushing contributors to move beyond this phase as it marks the end of their active involvement. 

\subsection{A Contributor's Journey through the Phases of Interaction with an OSS Project}

\Cref{fig:cross} offers an overview of the \textbf{CROSS} model, illustrating the phases of interaction between contributors and OSS projects, along with the forces at play and the mutual distance between them. However, it simplifies certain use cases, such as episodic contributors or progression through roles within the Membership phase, making them harder to interpret. To address this, we provide an alternate view in \Cref{fig:dev}, which focuses on a contributor's journey through the different phases. In this view, we can follow the contributor's move between two states—non-contributing and contributing. Initially, in the Discovery phase, the contributor is a `Project Outsider' in a non-contributing state. Transition to the Onboarding phase marks their entry into active contribution. During the Membership phase, contributors may either be engaged (handling tasks like issue resolution) or disengaged (contributing less actively). Engaged contributors may take on roles such as project maintainer or core developer. Contributors eventually move through the Offboarding phase and enter the Dropoff phase as Former Project Insiders, though they may return to Membership under certain conditions. A complete disengagement occurs when they cut all ties, but without confirmation from the contributor, this state is difficult to verify.

\section{Discussion}

\subsection{Archetypal Stakeholders}
The examples in this section represent archetypal stakeholders within the OSS ecosystem—individuals or groups who engage with OSS projects in distinct, predictable ways based on common patterns of behavior, characteristics, motivations, and interaction. Several archetypal OSS contributors have been identified by existing literature, e.g., episodic volunteers [\cite{barcomb2018uncovering}], core developers [\cite{mockus2000case}], one-time contributors [\cite{zhou2012make}], etc. Some archetypes, like episodic volunteers, are challenging to capture with models like the Onion model. Here, we describe how we can use the \textbf{CROSS} model to capture the behaviors of known archetypes of OSS contributors.

\begin{itemize}[wide=5pt,noitemsep,parsep=2pt,topsep=2pt]
    \item \textit{An outsider joins a project, becomes a maintainer, and eventually leaves} -- The contributor begins as a Project Outsider, discovering the OSS project through various channels such as recommendations, online searches, or community forums [\cite{VonKrogh2003}].
    During this initial phase, they explore project documentation, browse issue discussions, and may join mailing lists, engaging in anticipatory socialization.
    Once they decide to contribute actively, they transition into the Onboarding phase. Here, they navigate the learning curve by reviewing contribution guidelines, submitting pull requests, filing bug reports, and engaging with project maintainers for feedback and review.
    Upon making a successful contribution, they enter the Membership phase, where they contribute regularly, gain the community's trust, and may eventually be promoted to project maintainer.
    Over time, for various reasons, the contributor decides to disengage from the project and enter the offboarding phase. Ideally, this involves notifying the community, wrapping up ongoing tasks, and engaging in knowledge transfer to ensure that expertise is retained within the project.  
    Finally, they transition to the Dropoff phase, where they fully disengage and become a `former project insider'. While they no longer actively contribute, they may maintain occasional contact with the community or, in some cases, return to active contribution if circumstances or interests change. A real-world example of this can be seen for the ``node-pre-gyp''\footnote{\url{https://github.com/mapbox/node-pre-gyp/issues/657}, Accessed: 2024-09-04.} OSS project.
    \item \textit{Episodic Volunteers:} For episodic volunteers [\cite{barcomb2018uncovering}], the discovery and onboarding phase might work in a manner similar to the typical case. However, in the `Membership' phase, they participate by contributing sporadically based on their availability and interest. They may engage in specific tasks or projects within the community during certain periods but do not maintain continuous involvement like core or regular contributors. Depending on the frequency of their contributions, they might be considered to have offboarded and dropped off from the project between their spells of activity before returning and making contributions to the project again. This archetype of OSS contributors is hard to explain with the Onion model but the contributors' journey shown in \Cref{fig:dev} can succinctly capture this phenomenon.
    \item \textit{One Time Contributors:} A significant number of contributors to OSS projects are one-time contributors [\cite{zhou2012make}]. Although individual cases might be different, typically such contributors spend quite a lot of time in the `Discovery' phase until they find something they want to contribute to the project. Then they enter the `Onboarding' phase, engaging with the project members to get their contribution accepted. However, once their contribution is accepted, they spend almost no time in the `Membership' and `Offboarding' phases and move straight to the `Dropoff' phase. Once again, the contributors' journey shown in \Cref{fig:dev} helps us capture this using the \textbf{CROSS} model.
    \item \textit{Invited Contributors:} These are contributors to a project who are invited by the project maintainers (or a sponsoring organization in the case of sponsored projects) to address certain problems in the project. These contributors spend little to no time in the Discovery and Onboarding phases and assume membership in the project. They may or may not stick around, depending on the circumstances. If they do not, they undergo the Offboarding phase, handing over all responsibilities, and move on to the Dropoff phase. 
    \item \textit{Project Creator leaving Project:} As a core developer starting an OSS project, the project creators never go through the Discovery and Onboarding phases. However, they might spend some effort popularizing the project, establishing contribution guidelines, mentoring new contributors, and integrating their initial contributions into the project. Once the project takes off, the core developer assumes the role of a central figure within the project community. They actively contribute code, review pull requests, manage project milestones, and engage in discussions with other contributors. This phase represents a period of sustained involvement and leadership within the OSS project, where the core developer plays a pivotal role in guiding its development and direction. At some point, the core developer may decide to step back from active contribution or leadership roles, entering the Offboarding phase. During Offboarding, the core developer ideally delegates responsibilities to trusted contributors, conducts knowledge transfer sessions, and ensures a smooth transition to maintain project continuity. The Dropoff phase occurs when the core developer completely disengages from the project. Examples include the `s3funnel' and `urllib3' \footnote{https://shazow.net/posts/\\how-to-hand-over-an-open-source-project-to-a-new-maintainer/: Accessed 2024-09-04.} projects.
\end{itemize}


\subsection{CROSS Model: Potential Applications}
In addition to explaining the lifecycles of different types of contributors, the \textbf{CROSS} model offers valuable insights into the practical challenges faced by OSS communities. 
While there has been significant research on onboarding newcomers, many OSS communities still struggle with retaining contributors and helping them advance to more influential roles, such as maintainers or core developers. This issue is particularly relevant for contributors from underrepresented groups, who may face unique challenges, such as a lack of mentorship, limited access to resources, or feeling unwelcome in a project’s culture. The ``forces'' and ``retainers'' described in the \textbf{CROSS} model can help bring attention to these challenges, creating clearer paths for diverse contributors to move from peripheral to leadership positions in OSS communities.
Moreover, the \textbf{CROSS} model extends beyond the Onion model by addressing the offboarding phase, which is often overlooked in OSS research and practice. While the Onion model largely focuses on contributors who remain engaged, the \textbf{CROSS} model sheds light on the critical process of knowledge transfer when key contributors leave a project. This perspective is especially valuable for maintaining project sustainability and ensuring that expertise is not lost when core developers disengage. It also reveals the risks of long-term disengagement of core developers and the need for proactive strategies to address this. Although the \textbf{CROSS} model cannot directly solve these challenges, it provides OSS communities with a deeper understanding of the risks posed by poor offboarding practices and the importance of contingency planning.



\section{Limitations}
While the \textbf{CROSS} model provides a valuable theoretical framework for understanding the lifecycle of contributor engagement in OSS projects, it is important to acknowledge certain limitations. First, this paper is not intended as an empirical case study, and the level of empirical rigor typically associated with case studies, such as data-backed validation through mining and analyzing data from pull requests, forum discussions, or bug reports, etc. is beyond its scope. The \textbf{CROSS} model is designed to offer insights into typical contributor-project interactions, but it does not delve into detailed analysis from real-world OSS projects. As such, its role is to serve as a conceptual tool rather than an empirically validated solution.

Additionally, it should be noted that the \textbf{CROSS} model does not directly solve practical challenges within OSS communities. Instead, it provides a structured way to examine the stages of contributor-project interaction and reveals potential areas where interventions could be made. The model is intended to inform future research and guide project maintainers in understanding contributor dynamics but does not prescribe specific solutions for issues like knowledge transfer or offboarding.

Finally, while the model addresses key phases like offboarding and role transitions, there may be other factors in OSS contributor behavior that are not fully captured. For example, external factors such as contributor availability, project governance structures, or even broader community trends may influence engagement in ways that fall outside the scope of the model. These factors represent opportunities for further refinement of the model through empirical research.

\section{Conclusion}

The \textbf{CROSS} model provides a comprehensive framework for understanding the full lifecycle of contributor engagement in OSS projects, addressing key phases such as onboarding, membership, offboarding, and dropoff. By bringing focus to contributor offboarding and disengagement from OSS projects, it highlights the critical importance of knowledge transfer and project continuity when key contributors leave.
By encompassing diverse contributor behaviors, such as those of episodic and one-time contributors, it offers a more nuanced understanding of the OSS ecosystem compared to existing models like the Onion model. Through its focus on drivers and retainers, the \textbf{CROSS} model explains how contributors transition between different phases of engagement or disengage altogether.

While the \textbf{CROSS} model is primarily a theoretical framework, it sets the stage for new research in several possible directions, including but not limited to OSS project sustainability, health metrics that capture long-term community resilience, structured exit processes, ensuring knowledge retention, and developing strategies to re-engage former contributors, who might otherwise be lost to the project. Its lifecycle-based approach captures both common and less-explored scenarios like episodic re-engagement and paves the way for more sustainable, inclusive, and resilient OSS ecosystems.

\section*{Acknowledgement}
The work was supported, in part, by Science Foundation Ireland grant 13/RC/2094 P2 to Lero.
\addtolength{\textheight}{-.2cm} 

\printbibliography

\end{document}